\documentclass{aa}
\usepackage{graphicx}
\begin{document}
   \title{Environmental effects in galaxies}

   \subtitle{Molecular Gas, Star Formation, and Activity{\thanks{Based on 
   observations at the European Southern Observatory at the 
   15m Swedish ESO Submillimetre telescope, SEST, and at the 
    the 1.52m telescope which is operated under the ESO-ON agreement.}}}

   \author{Duilia F. de Mello
          \inst{1}
           Tommy Wiklind\inst{1}
	   \and
	   Marcio A. G. Maia\inst{2}
          }

   \offprints{D. de Mello}

   \institute{Onsala Space Observatory,
              43992 Onsala, Sweden\\
              email: duilia@oso.chalmers.se, tommy@oso.chalmers.se 
	     \and
      Observat\'orio Nacional, Rua Gal. Jos\'e Cristino 77, RJ 20921, Brazil\\
      email: maia@on.br}      
       
   \date{Received; accepted}

   \abstract{
In order to study whether there is any correlation between nuclear activities, 
gas content, and the environment where galaxies 
reside, we have obtained optical and millimetric spectra for a well-defined sample of 
intermediate Hubble type spirals in dense environments and in the field. 
We found that these spirals in dense environments have on average: less 
molecular gas per blue luminosity, higher atomic gas fraction, lower current 
star formation rate, and the same star formation efficiency as field galaxies.
Although none of these results stand out as a single strong diagnostic, given their 
statistical significance, taken together they indicate a trend for diminished 
gas content and star formation activity in galaxies in high density environments.
Our results suggest that galaxies in dense environments have either (i) consumed 
their molecular gas via star formation in the past or (ii) that dense environments 
leads to an inhibition of molecular gas from atomic phase. The similarities in star formation 
efficiency of the dense environments and field galaxies suggest that the physical 
processes controling the formation of stars from the molecular gas are local rather than global.
We also found that star formation rate per blue luminosity increases linearly as the total amount of 
gas increases in LINERs. This result, based on a small sample, suggests that LINERs 
are powered by star formation rather than an AGN.
   }
\maketitle

\section{Introduction}

The importance of interactions for triggering activity
in galaxies has been extensively explored in the past
few decades (e.g. Larson \& Tinsley 1978, Dahari 1984, Kennicutt \& Keel 1984, 
Keel et al. 1985, Mihos \& Hernquist 1994, Liu \& Kennicutt 1995, 
Keel 1996). There is no doubt that the environment where galaxies reside plays 
a decisive role in galaxy evolution. Nevertheless, there are a few key questions which 
are still being debated. For instance, the environmental influences on the star 
formation properties of clusters of galaxies is far from clear-cut. Although 
the molecular gas properties of strongly HI deficient spirals in the Virgo cluster 
is similar to field spirals (Kenney \& Young 1988), the average star formation 
activity among them are lower than for a sample of field spirals (Kennicutt 1983).
This latter effect is, however, not seen in the Coma, Cancer, and A1367
clusters, where star formation activity appears to be enhanced with respect
to field spirals (Kennicutt et al. 1984). The Coma spirals are similar to
those in the Virgo cluster: deficient in atomic gas, while the molecular gas
properties are the same as for field spirals (Casoli et al. 1991, 1996; Gerin
\& Casoli 1994). In the Fornax cluster, Horellou et al. (1995) found no evidence for HI
deficiency, but an unusual low fraction of molecular gas.  For galaxies in loose groups,  
Maia et al. (1998) find only weak evidence for HI depletion in the 
early-type spirals. 

In compact groups of galaxies the environmental role is also an open issue. 
Compact groups of galaxies are longlived entities with a
space density of galaxies higher than in clusters (see Hickson 1997 for a 
review). Triggering of star formation through gravitational interaction 
should therefore be even more important in this environment, but the results indicate differently. 
Even though compact groups present a high fraction of
distorted galaxies (Mendes de Oliveira \& Hickson 1994), they do not show an 
enhancement in far-infrared (FIR) emission (Sulentic \& de Mello Raba\c ca 1993, 
Allam et al. 1996), they have low HI content (Williams \& Rood 1987, Huchtmeier 1997), and a 
normal CO content (Boselli et al. 1996, Leon et al. 1998).

In pairs of galaxies, the scenario is different. The CO and far--infrared
luminosities, normalized with either the size of the respective galaxy or
L$_{\rm B}$, is enhanced (Combes et al. 1994). However, whereas the star 
formation efficiency (SFE = star formation rate per mass unit of molecular gas) is 
higher in pairs which are strongly interacting/merging, 
the average SFE of the whole sample of pairs is similar to normal field spirals.  
A possible interpretation is that gravitational interactions do not increase 
the SFE, but increases the amount of star forming gas, 
possibly through infall of new material. 

It has been shown that strong gravitational interactions between galaxies can 
enhance the star formation rate (SFR) (e.g. Liu \& Kennicutt 1995). It has also been proposed that it is the 
near environment that mostly affects the evolution of galaxies (Szomoru et al. 1996).
However, the connection between the environment, nuclear activities and total gas content 
has not been able to explain which variables are important in deciding how efficient 
stars are formed. Are galaxies in dense environments more efficient in forming stars or do they
have more fuel? How is the star formation efficiency correlated with the environment 
where galaxies reside? Is the near environment that mostly affects the evolution of
galaxies. An essential step towards answering these 
open questions is to compare the properties of galaxies in dense environments and in the 
field.

In this work we present the analysis of the data shown in de Mello et al. 
(2001, hereafter Paper I) which is a database of molecular and optical spectra 
of galaxies in dense regions of the Southern sky and in the field. 

This paper is organized as follows. 
Section 2 presents the diagnostics used in this work. 
Discussion is presented in Section 3
and a summary of the main results and conclusions are presented in Section 4.


\section{Diagnostics}

Optical and millimetric data have been obtained with the ESO1.52m and the SEST
15m radio telescope in la Silla, Chile and we refer to Paper I for further
details. Table~\ref{results} lists the data as follows. 
Column (1): designation in the ESO-Uppsala catalog (Lauberts \& Valentijn 1989, hereafter LV89); 
column (2): type of sample (control sample=CS and high 
density sample=HDS) and morphological type (LV89) 
1=Sa, 2=Sa-b, 3=Sb, 4=Sb-c, 5=S..., 6=Sc, Sc-d, 7=S../Irr, 8=Sd; 
column (3): velocity derived from central CO (1-0) profiles in kms$^{-1}$; 
column (4): distance in Mpc corrected for the Virgocentric flow according to 
model 3.1 in Aaronson  et al. (1982); 
column (5): blue luminosity in L$_{\odot}$ derived from B$_{\rm T}$ magnitude
 taken from RC3; 
column (6): HI masses in M$_{\odot}$ derived using the relation 
M$_{\rm HI}$ = 2.36 $\times$ 10$^{5}$ $\times$ D$^{2}$ $\times$ F(HI),
where M$_{\rm HI}$ is the HI mass in M$_{\odot}$, D is the distance in Mpc, and F(HI) 
is the HI flux in Jy kms$^{-1}$ taken from the NASA/IPAC Extragalactic Database (NED) (blank = no HI data available); 
column (7): Far-Infrared luminosity in L$_{\odot}$ calculated from 
L$_{\rm FIR}$ = 5.9$\times$10$^{5}$D$^{2}$(2.58$\times$F$_{60}$+F$_{100}$) where D is the distance in Mpc
and F are IRAS fluxes at 60 and 100 $\mu$m (Moshir et al. 1990); 
column (8): H$_{\rm 2}$ masses in M$_{\odot}$ estimated from the velocity integrated emission, using a 
N$_{\rm H_{2}}$/I$_{\rm CO}$ conversion ratio of 3$\times$10$^{20}$
cm$^{-2}$ (K kms$^{-1}$); 
column (9): dust temperature in K calculated as described in Sect.~2.3; 
column (10): dust masses in M$_{\odot}$ calculated as described in Sect.~2.3; and column (11): type of activity (L = LINERs, HII = HII region, blank = no optical data) classified as described in
Sect.~2.2. A Hubble constant value of 75 kms$^{-1}$Mpc$^{-1}$ was adopted in all calculations.

\begin{table*}
\caption[]{The Data}
\label{results}
\begin{tabular}{llllrrrrrcc}
\hline
\multicolumn{1}{c}{ESO-LV}   &
\multicolumn{1}{c}{Sample \&} &
\multicolumn{1}{c}{V$_{\rm CO}$} &
\multicolumn{1}{c}{Dist.}&
\multicolumn{1}{c}{log L$_{\rm B}$} &
\multicolumn{1}{c}{log M$_{\rm HI}$} &
\multicolumn{1}{c}{L$_{\rm FIR}$$\times$$10^{9}$} &
\multicolumn{1}{c}{M$_{\rm H_2}$$\times$$10^{9}$} &
\multicolumn{1}{c}{T$_{\rm dust}$} &
\multicolumn{1}{c}{M$_{\rm dust}$$\times$$10^{6}$} &
\multicolumn{1}{c}{Type of}\\
\multicolumn{1}{c}{Name} &
\multicolumn{1}{c}{Morph.} &
\multicolumn{1}{c}{kms$^{-1}$} &
\multicolumn{1}{c}{Mpc} &
\multicolumn{1}{c}{L$_{\odot}$} &
\multicolumn{1}{c}{M$_{\odot}$} &
\multicolumn{1}{c}{L$_{\odot}$} &
\multicolumn{1}{c}{M$_{\odot}$} &
\multicolumn{1}{c}{K}&
\multicolumn{1}{c}{M$_{\odot}$}&
\multicolumn{1}{c}{Activity}\\
\multicolumn{1}{c}{(1)} &
\multicolumn{1}{c}{(2)} &
\multicolumn{1}{c}{(3)} &
\multicolumn{1}{c}{(4)} &
\multicolumn{1}{c}{(5)} &
\multicolumn{1}{c}{(6)} &
\multicolumn{1}{c}{(7)} &
\multicolumn{1}{c}{(8)} &
\multicolumn{1}{c}{(9)} &
\multicolumn{1}{c}{(10)}&
\multicolumn{1}{c}{(11)}\\
\hline
0310050 & CS  3.5&   4714 & 59.2 &10.11 &      &13.60 $\pm$ 0.39 & 3.30 $\pm$ 0.23 & 28.64 $\pm$ 0.49 & 9.98 $\pm$ 0.95 & HII\\
1060120 & CS  6  &   4154 & 52.0 & 9.97 &      & 6.90 $\pm$ 0.39 & 1.93 $\pm$ 0.19 & 29.45 $\pm$ 1.03 & 4.24 $\pm$ 0.81 & HII\\
1080130 & HDS 3.5&   2941 & 35.8 & 9.78 &      & 2.67 $\pm$ 0.15 & 0.81 $\pm$ 0.07 & 28.91 $\pm$ 1.00 & 1.85 $\pm$ 0.35 & HII\\
1080200 & CS  3.9&   1719 & 19.9 & 9.37 & 9.63 & 4.45 $\pm$ 0.20 & 0.65 $\pm$ 0.03 & 28.71 $\pm$ 0.82 & 3.22 $\pm$ 0.49 & \\
1190060 & HDS 7.5&   1256 & 14.4 & 9.48 & 8.91 & 1.66 $\pm$ 0.05 & 0.11 $\pm$ 0.01 & 32.62 $\pm$ 0.72 & 0.55 $\pm$ 0.06 & \\
1190190 & HDS 5  &   1527 & 18.0 & 9.94 & 9.14 & 2.52 $\pm$ 0.07 & 0.42 $\pm$ 0.02 & 28.85 $\pm$ 0.50 & 1.77 $\pm$ 0.17 & HII\\
1420500 & CS  5  &   2135 & 25.9 &10.10 & 9.98 & 4.40 $\pm$ 0.09 & 0.59 $\pm$ 0.04 & 28.78 $\pm$ 0.44 & 3.12 $\pm$ 0.24 & L \\
1460090 & CS  5  &   1652 & 19.0 &10.131& 9.47 &10.40 $\pm$ 0.34 & 1.10 $\pm$ 0.06 & 30.88 $\pm$ 0.70 & 4.75 $\pm$ 0.51 & HII\\
1570050 & HDS 5.5&   1326 & 15.2 & 9.30 & 8.84 & 0.39 $\pm$ 0.02 & 0.10 $\pm$ 0.01 & 29.27 $\pm$ 0.96 & 0.25 $\pm$ 0.05 & HII\\
1890070 & CS  4.0&   3006 & 36.7 &10.44 &      & 9.15 $\pm$ 0.36 & 1.48 $\pm$ 0.09 & 31.74 $\pm$ 0.83 & 3.54 $\pm$ 0.46 & \\
2010220 & CS  5  &   3990 & 50.1 & 9.70 & 9.90 & 3.53 $\pm$ 0.23 & 0.98 $\pm$ 0.07 & 27.81 $\pm$ 1.03 & 3.14 $\pm$ 0.67 & HII\\
2030180 & CS  4  &   4123 & 52.2 &10.27 &      &25.39 $\pm$ 1.17 & 3.30 $\pm$ 0.16 & 32.28 $\pm$ 1.02 & 8.88 $\pm$ 1.41 & HII\\
2340160 & HDS 5  &   5218 & 66.4 &10.01 &      & 3.72 $\pm$ 0.50 & 0.94 $\pm$ 0.06 & 32.22 $\pm$ 3.00 & 1.32 $\pm$ 0.63 & HII\\
2350550 & HDS 5  &   5098 & 64.6 &10.73 &      & 9.92 $\pm$ 1.11 & 2.04 $\pm$ 0.12 & 24.89 $\pm$ 1.22 & 8.61 $\pm$ 6.00 &  L\\
2350570 & HDS 4  &   5069 & 64.2 &10.03 &      &11.08 $\pm$ 1.30 & 3.45 $\pm$ 0.13 & 23.83 $\pm$ 1.23 & 8.20 $\pm$ 9.77 & L \\
2370020 & CS  4.5&   5173 & 65.3 &10.58 &10.21 &13.72 $\pm$ 0.70 & 4.90 $\pm$ 0.18 & 27.44 $\pm$ 0.78 & 3.31 $\pm$ 2.20 & L\\
2400110 & HDS 4.8&   2890 & 34.9 &10.00 &10.26 & 5.81 $\pm$ 0.31 & 1.65 $\pm$ 0.05 & 25.20 $\pm$ 0.93 & 9.98 $\pm$ 2.16 & L \\
2400130 & HDS 3  &   3284 & 40.1 & 9.80 &      & 5.17 $\pm$ 0.31 & 1.08 $\pm$ 0.06 & 27.44 $\pm$ 0.85 & 5.02 $\pm$ 0.94 & \\
2850080 & HDS 4  &   2838 & 35.3 &10.63 &10.34 & 4.45 $\pm$ 0.23 & 0.64 $\pm$ 0.07 & 26.69 $\pm$ 0.74 & 5.18 $\pm$ 0.86 & L \\
2860820 & HDS 5  &   4958 & 62.8 & 9.98 &      & 4.42 $\pm$ 0.51 & 1.66 $\pm$ 0.09 & 30.44 $\pm$ 2.23 & 2.21 $\pm$ 0.87 & HII \\
2880260 & HDS 5  &   2383 & 28.8 & 9.79 & 9.57 & 1.42 $\pm$ 0.10 & 0.32 $\pm$ 0.02 & 29.40 $\pm$ 1.29 & 0.88 $\pm$ 0.21 & L \\
2960380 & CS  4  &   3645 & 45.1 & 9.90 &      & 3.73 $\pm$ 0.33 & 0.44 $\pm$ 0.06 & 29.33 $\pm$ 1.69 & 2.35 $\pm$ 0.72 & HII \\
3050140 & CS  5  &   4761 & 61.1 &10.11 & 9.78 & 4.31 $\pm$ 0.55 & 2.31 $\pm$ 0.10 & 32.15 $\pm$ 2.80 & 1.55 $\pm$ 0.71 & HII \\
3470340 & HDS 3  &   1671 & 19.3 & 9.92 & 9.70 & 7.85 $\pm$ 0.79 & 2.27 $\pm$ 0.06 & 28.00 $\pm$ 1.54 & 6.67 $\pm$ 2.14 & \\
3500140 & CS  6  &   3400 & 42.0 &10.09 & 9.73 & 3.30 $\pm$ 0.25 & 1.23 $\pm$ 0.04 & 30.04 $\pm$ 1.49 & 1.79 $\pm$ 0.47 & HII\\
3520530 & HDS 3  &   3874 & 48.2 &10.27 & 9.20 &21.89 $\pm$ 0.93 & 6.55 $\pm$ 0.32 & 30.53 $\pm$ 0.83 & 0.75 $\pm$ 1.54 & \\
3550260 & CS  4  &   1985 & 23.8 & 9.42 & 8.82 & 0.95 $\pm$ 0.07 & 0.15 $\pm$ 0.02 & 29.75 $\pm$ 1.46 & 0.55 $\pm$ 0.14 & \\
3550300 & CS  4  &   4448 & 56.1 &10.25 & 9.82 &10.05 $\pm$ 0.43 & 3.33 $\pm$ 0.36 & 29.04 $\pm$ 0.75 & 6.75 $\pm$ 0.96 & L\\
3570190 & HDS 5  &   1789 & 21.4 & 9.83 & 9.43 & 1.52 $\pm$ 0.06 & 0.41 $\pm$ 0.03 & 28.48 $\pm$ 0.67 & 1.15 $\pm$ 0.15 & HII\\
4050180 & CS  1  &   3375 & 41.9 &10.27 & 8.87 &10.67 $\pm$ 0.68 & 3.52 $\pm$ 0.17 & 33.19 $\pm$ 1.50 & 3.17 $\pm$ 0.70 & \\
4060250 & HDS 5  &   1470 & 16.9 & 9.98 & 9.26 & 4.42 $\pm$ 0.20 & 2.04 $\pm$ 0.05 & 29.31 $\pm$ 0.83 & 2.80 $\pm$ 0.43 & \\
4060330 & HDS 6  &   1922 & 22.8 & 9.71 & 9.72 & 5.01 $\pm$ 0.21 & 0.42 $\pm$ 0.02 & 31.61 $\pm$ 0.90 & 1.99 $\pm$ 0.29 & HII\\
4070140 & CS 5   &   2761 & 33.6 & 9.85 & 9.54 & 3.54 $\pm$ 0.23 & 0.78 $\pm$ 0.04 & 31.34 $\pm$ 1.33 & 1.48 $\pm$ 0.33 & HII\\
4190030 & CS 4   &   4146 & 52.8 &10.20 & 9.82 &11.19 $\pm$ 0.36 & 1.10 $\pm$ 0.09 & 32.48 $\pm$ 0.71 & 3.77 $\pm$ 0.42 & HII\\
4200030 & CS 5   &   4093 & 52.2 &10.22 & 9.83 & 6.41 $\pm$ 0.41 & 2.02 $\pm$ 0.14 & 30.37 $\pm$ 1.24 & 3.25 $\pm$ 0.70 & HII\\
4710200 & CS 4.5 &   3017 & 37.0 &10.30 &10.17 &12.49 $\pm$ 0.53 & 1.95 $\pm$ 0.13 & 30.08 $\pm$ 0.81 & 6.72 $\pm$ 0.96 & HII\\
4780060 & CS 4   &   5401 & 68.9 &10.58 & 9.95 &51.12 $\pm$ 2.91 & 10.86 $\pm$ 0.4 & 32.26 $\pm$ 1.26 & 7.96 $\pm$ 3.55 & HII\\
4820430 & CS 4   &   4073 & 51.9 &10.17 & 9.74 & 6.57 $\pm$ 0.33 & 2.35 $\pm$ 0.21 & 26.86 $\pm$ 0.81 & 7.33 $\pm$ 1.27 & HII\\
4840250 & CS 2   &   4128 & 53.0 &10.13 &      &16.54 $\pm$ 0.64 & 2.65 $\pm$ 0.22 & 32.17 $\pm$ 0.85 & 5.90 $\pm$ 0.79 & \\
5320090 & CS 5   &   2582 & 32.0 & 9.91 & 9.31 & 4.22 $\pm$ 0.20 & 0.41 $\pm$ 0.05 & 31.61 $\pm$ 0.98 & 1.67 $\pm$ 0.26 & HII\\
5390050 & CS 5   &   3158 & 39.4 & 9.98 & 8.99 & 5.03 $\pm$ 0.29 & 1.99 $\pm$ 0.13 & 30.51 $\pm$ 1.11 & 2.48 $\pm$ 0.47 & HII\\
5450100 & HDS 5  &   1715 & 20.9 & 9.55 & 9.07 & 3.21 $\pm$ 0.14 & 0.15 $\pm$ 0.01 & 34.52 $\pm$ 1.08 & 0.76 $\pm$ 0.11 & HII\\
5450110 & HDS 5  &   1456 & 17.5 &10.35 & 9.60 &14.78 $\pm$ 0.72 & 2.15 $\pm$ 0.07 & 30.79 $\pm$ 0.91 & 6.88 $\pm$ 1.14 & \\
5480070 & HDS 3.5&   1557 & 19.2 & 9.87 & 9.38 & 1.05 $\pm$ 0.05 & 0.17 $\pm$ 0.01 & 26.19 $\pm$ 0.80 & 1.39 $\pm$ 0.25 & \\
5480310 & HDS 3  &   1531 & 18.9 & 9.79 & 8.60 & 3.17 $\pm$ 0.13 & 0.65 $\pm$ 0.03 & 28.87 $\pm$ 0.79 & 2.21 $\pm$ 0.31 & L\\
5480380 & HDS 6  &   1874 & 23.3 &10.03 & 9.23 &10.56 $\pm$ 0.31 & 0.30 $\pm$ 0.02 & 41.99 $\pm$ 1.08 & 0.87 $\pm$ 0.09 & HII\\
6010040 & CS 4.6 &   5219 & 66.8 &10.01 & 9.73 & 3.85 $\pm$ 0.56 & 1.17 $\pm$ 0.09 & 28.33 $\pm$ 2.74 & 3.03 $\pm$ 1.57 & HII\\
\hline
\end{tabular}
\noindent CS=control sample and HDS=high density sample; morphological types are 
1=Sa, 2=Sa-b, 3=Sb, 4=Sb-c, 5=S..., 6=Sc, Sc-d, 7=S../Irr, 8=Sd. Column(11): HII=activity typical of HII regions, L=LINERs, blank means no optical 
data. $^\dagger$ M$_{\rm H_{2}}$ of 5 points along the major axis.
$^\ddagger$ M$_{\rm H_{2}}$ of 7 points along the major axis.
\end{table*}

\begin{table*}
\caption[]{Diagnostic Quantities}
\scriptsize
\label{tabdiag}
\begin{tabular}{lccccccc}
\hline
\multicolumn{1}{c}{Sample$^\dagger$} &
\multicolumn{1}{c}{log (L$_{\rm FIR}$/M$_{\rm H_2}$)}&
\multicolumn{1}{c}{log (M$_{\rm H_2}$/L$_{\rm B}$)}&
\multicolumn{1}{c}{log (L$_{\rm FIR}$/L$_{\rm B}$)}&
\multicolumn{1}{c}{log (M$_{\rm H_2}$+M$_{\rm HI}$/L$_{\rm B}$)} &
\multicolumn{1}{c}{log (L$_{\rm FIR}$/M$_{\rm H_2}$+M$_{\rm HI}$)}&
\multicolumn{1}{c}{T$_{\rm D}$}&
\multicolumn{1}{c}{log (M$_{\rm D}$/L$_{\rm B}$)}\\
\multicolumn{1}{c} {}&
\multicolumn{1}{c} {L$_{\odot}$/M$_{\odot}$}&
\multicolumn{1}{c} {M$_{\odot}$/L$_{\odot}$}&
\multicolumn{1}{c} {L$_{\odot}$/M$_{\odot}$}&
\multicolumn{1}{c} {M$_{\odot}$/L$_{\odot}$}&
\multicolumn{1}{c} {L$_{\odot}$/M$_{\odot}$}&
\multicolumn{1}{c} {K}&
\multicolumn{1}{c} {M$_{\odot}$/L$_{\odot}$}\\
\hline
HDS        & 0.74 $\pm$ 0.31 & -1.09 $\pm$ 0.39 & -0.35 $\pm$ 0.32& -0.35 $\pm$ 0.29& -0.01 $\pm$ 0.42&
31.3 $\pm$ 2.8 & -3.65 $\pm$ 0.27\\
CS         & 0.67 $\pm$ 0.21 & -0.91 $\pm$ 0.24 & -0.24 $\pm$ 0.22& -0.27 $\pm$ 0.25&  0.00 $\pm$ 0.26&
30.1 $\pm$ 2.4 & -3.48 $\pm$ 0.30\\
Pairs      & 0.91 $\pm$ 0.43 & -0.57 $\pm$ 0.45 &  0.33 $\pm$ 0.48& -0.47 $\pm$ 0.31&  1.04 $\pm$ 0.37&
34.9 $\pm$ 6.0 & -3.37 $\pm$ 0.40\\
HCG        & 0.39 $\pm$ 0.33 & -0.61 $\pm$ 0.39 & -0.16 $\pm$ 0.45& -0.42 $\pm$ 0.22& -0.02 $\pm$ 0.40&
33.1 $\pm$ 5.7 & -3.42 $\pm$ 0.36\\
HCG$_{\rm int.type}$& 0.37 $\pm$ 0.40 & -0.66 $\pm$ 0.35 & -0.29 $\pm$ 0.49& -0.14 $\pm$ 0.31& -0.12 $\pm$ 0.40&
32.3 $\pm$ 5.6 & -3.41 $\pm$ 0.32\\
Starbursts & 1.24 $\pm$ 0.39 & -0.61 $\pm$ 0.43 &  0.63 $\pm$ 0.43& -0.36 $\pm$ 0.40&  0.91 $\pm$ 0.39&
40.4 $\pm$ 6.2 & -3.27 $\pm$ 0.34\\
Clusters   & 0.77 $\pm$ 0.37 & -1.08 $\pm$ 0.36 & -0.31 $\pm$ 0.40& -0.73 $\pm$ 0.34&  0.42 $\pm$ 0.35&
33.2 $\pm$ 4.7 & -3.78 $\pm$ 0.30\\         
\hline
\end{tabular}
\noindent $^\dagger$ HDS is our high density sample, CS is our control sample 
of isolated galaxies, Pairs, Hickson Compact Groups (HCG), Starbursts and 
Clusters are from Leon et al. (1998). HCG$_{\rm int.type}$ are HCG galaxies of types 
Sb, Sbc, and Sc in Leon et al (1998).
\end{table*}  

\begin{figure*}
\centering
\includegraphics[width=\textwidth]{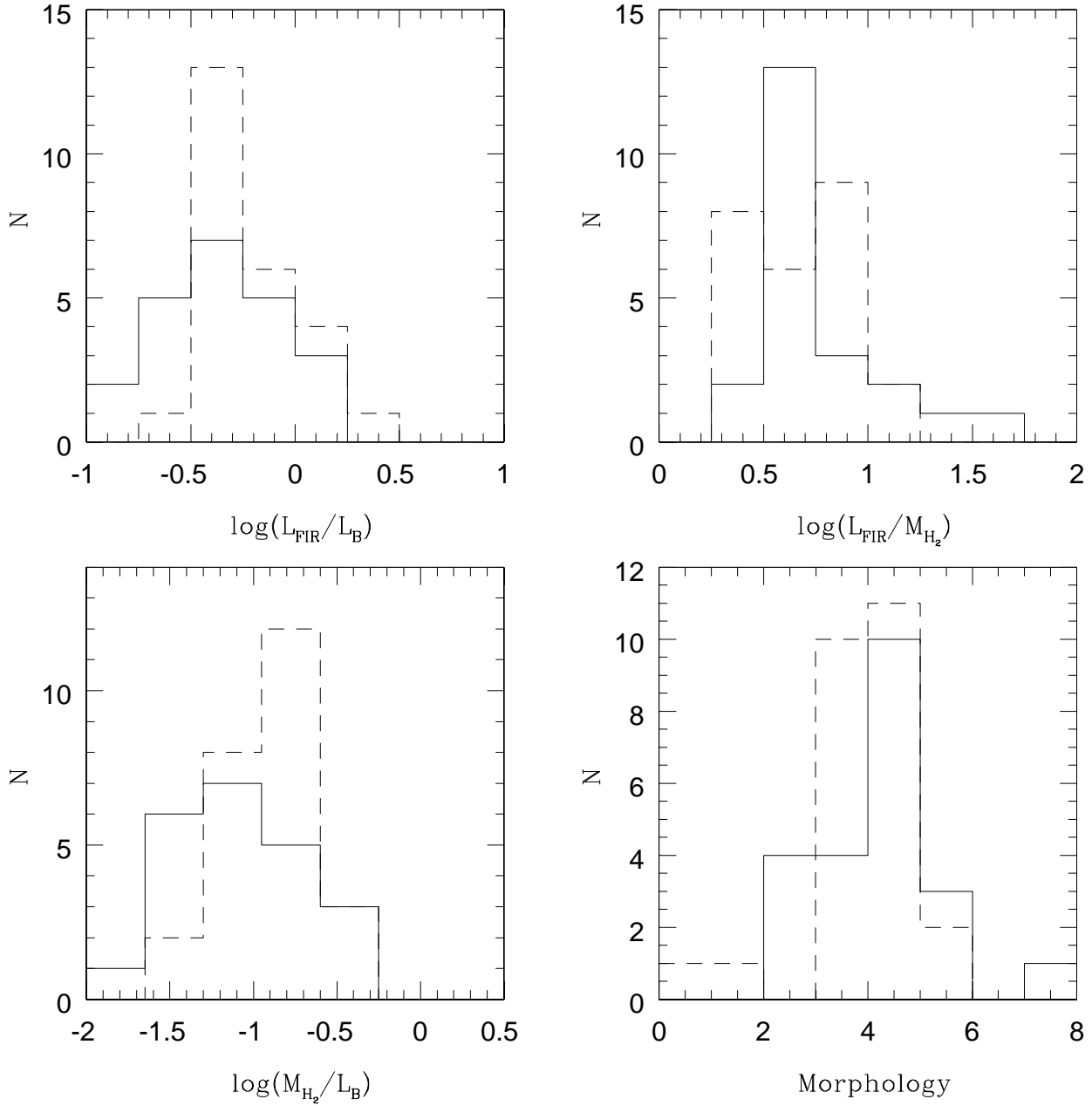}
\caption{Upper left panel: Distribution of FIR luminosity normalized by blue luminosity;
Upper right panel: Distribution of FIR luminosity normalized by molecular gas;
Lower left panel: Distribution of molecular gas normalized by blue luminosity;
Lower right panel: Distribution of morphological types. Morphological types are: 
1=Sa, 2=Sa-b, 3=Sb, 4=Sb-c, 5=S..., 6=Sc, Sc-d, 7=S../Irr, 8=Sd.
Full line is for the HDS and dashed line
is for the CS.  Luminosities are in L$_{\odot}$, mass in M$_{\odot}$.}
\label{MS10623f1}
\end{figure*}

\begin{figure*}
\centering
\includegraphics[width=\textwidth]{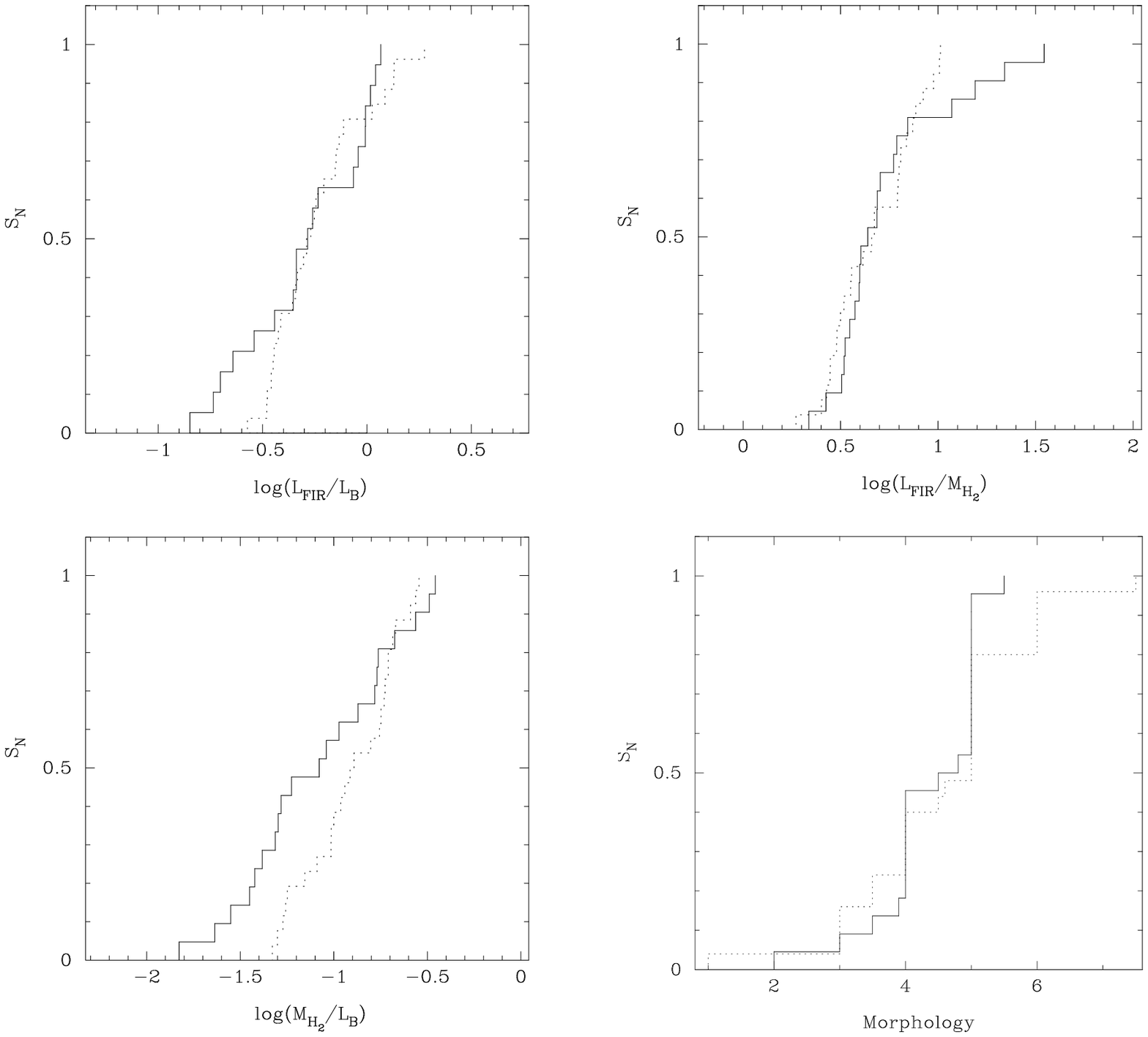}
\caption{Cumulative probability distribution of the same parameters of Fig. 1.}
\label{MS10623f2}
\end{figure*}

The diagnostics used in our search for environmental effects 
are summarized in Table~\ref{tabdiag} and Table~\ref{coeffs}. 
Due to the presence of galaxies with higher L$_{\rm B}$ in the CS 
(a distance bias in our subsample), masses and luminosities were normalized 
by L$_{\rm B}$. Given our morphological selection criteria, we assumed that 
the mass/L$_{\rm B}$ ratio is approximately the same for our galaxies 
and L$_{\rm B}$ is thus a measure of the total mass (e.g. Roberts \& Haynes 1994). 
In  Fig. 3a of Paper I we have investigated whether the bias in blue luminosity present in our 
subsample may cause a bias in our analysis. The correlation found for HDS and CS when we
plotted M$_{\rm H_{2}}$/L$_{\rm B}$ as a function of L$_{\rm B}$ is very similar suggesting 
no evident bias.

We included in Table~\ref{tabdiag} the average values 
given by Leon et al. (1998) for pairs of galaxies, Hickson Compact Groups,
starbursts and clusters. We have also included the average values from 
Leon et al. for galaxies in compact groups using the same morphological
criterion we used in our selection (Sb, Sbc, and Sc).

The distributions of the diagnostics are shown in Fig.~\ref{MS10623f1}. 
The cumulative distributions are shown in Fig~\ref{MS10623f2}.
The distribution of morphological types for the HDS and the CS were
also included in order to verify how similar the two samples were in terms of
morphology. This is a very important aspect to be considered in this type 
of analysis since morphological appearance is directly correlated with 
general properties of galaxies. We refer to Section 2.4 of Paper I for more
details on our morphology selection and to Roberts \& Haynes (1994) for a review
on physical parameters along the Hubble sequence.

The significance of the small difference between the mean values of the HDS and the CS
was assessed using the Student-t test for unequal variances. 
The Kolmogorov-Smirnov statistics (KS, hereafter) was used to assess the significance level of 
the difference between the cumulative distributions. Table~\ref{stat} shows: column (1) the value
of KS (same as $D$ in Press et al. 1989), which is the greatest distance between two 
cumulative distributions in the KS statistics; column (2) the significance level KS$_{\rm Pb}$; 
column (3) the Student-t coefficient T for unequal variances; and column (4) the significance level T$_{\rm Pb}$. 
Small values of T$_{\rm Pb}$ indicate that the distributions have 
significantly different means. Small values of KS$_{\rm Pb}$ indicate that the 
cumulative distribution of the HDS is significantly different from that of the CS.
The main results are:

\begin{itemize}

\item The HDS and the CS have similar morphology distribution. 

\item The HDS has on average lower M$_{\rm H_{2}}$/L$_{\rm B}$ than the CS.
M$_{\rm H_{2}}$/L$_{\rm B}$ distributions are significantly different.
Their means differ at the 93\% level. 

Therefore, {\it HDS spirals have overall less molecular gas 
per blue luminosity than spirals in the field}.

\item The HDS has on average lower L$_{\rm FIR}$/L$_{\rm B}$ than the CS. 
L$_{\rm FIR}$/L$_{\rm B}$ distributions are different at the 67\% level.
Their means differ at the 84\%  level. 

Therefore, {\it the HDS spirals have, on average, lower L$_{\rm FIR}$/L$_{\rm B}$ 
than the CS}.

\item The L$_{\rm FIR}$/M$_{\rm H_2}$ ratio can be interpreted as a measure of the 
star formation efficiency (star formation rate per unit mass of molecular gas). 
The HDS has on average higher L$_{\rm FIR}$/M$_{\rm H_2}$ than the CS.
However, the level of significance given by the Stundent-t test and KS statistics 
is only at the 60\% and 57\% level. 

Therefore, {\it the star formation efficiency in the HDS is not statistically different than
in the CS}.

\end{itemize}  
\subsection{Comparing with other samples}

We showed in Paper I that our subsamples of the HDS and the CS have general
properties, such as FIR luminosity, blue luminosity, and molecular gas content,  
very similar to other galaxies such as normal spiral galaxies (Young et al. 1989, 
Braine et al. 1993), ultraluminous FIR galaxies (Sanders et al. 1991), 
and galaxies in the Coma and Fornax clusters (Casoli et al. 1991 and Horellou et al. 1995); 
i.e. they are not a separate class of objects.

However, our M$_{\rm H_{2}}$/L$_{\rm B}$ average value 
(log(M$_{\rm H_{2}}$/L$_{\rm B}$)= -0.91 $\pm$ 0.24 for the CS) 
is lower than the classical value from Young \& Knezek (1989) 
(log(M$_{\rm H_{2}}$/L$_{\rm B}$) $\sim$ --0.77 $\pm$ 0.05 M$_{\odot}$/L$_{\odot}$).
The main reason for this disagreement is that Young \& Knezek sample of spirals 
was FIR-selected, it is biased towards higher L$_{\rm FIR}$ and therefore towards 
higher M$_{\rm H_{2}}$. A sample like CS which is not selected with any FIR limit, is more realistic 
for the field. 

Another sample of isolated galaxies such as the one used in Leon et 
al. (1998) shows a high dispersion of log(M$_{\rm H_{2}}$/L$_{\rm B}$) 
$\sim$~--0.78 $\pm$ 0.58 M$_{\odot}$/L$_{\odot}$ which demonstrates that their sample 
is not very homogeneous in terms of molecular gas content. This is 
due to the fact that the sample of Leon et al. (1998) includes all morphological types,
whereas our sample has only galaxies Sb, Sbc, and Sc. This is also seen 
when one compares Hickson Compact Groups of all types with HCG later 
than Sa and earlier than Sd given in Table~\ref{tabdiag}.

\begin{table}
\caption[]{Correlation Coefficients}
\label{coeffs}
\begin{tabular}{lcc}
\hline
\multicolumn{1}{c}{Diagnostic} &
\multicolumn{1}{c}{r} &
\multicolumn{1}{c}{r} \\
\multicolumn{1}{c}{} &
\multicolumn{1}{c}{HDS} &
\multicolumn{1}{c}{CS} \\
\hline
L$_{\rm FIR}$ $\times$ M$_{\rm H_2}$ & 0.80 & 0.84 \\
L$_{\rm B}$ $\times$ M$_{\rm H_2}$ & 0.64 & 0.78 \\
L$_{\rm FIR}$/L$_{\rm B}$ $\times$ M$_{\rm H_2}$/L$_{\rm B}$ & 0.65 & 0.59 \\
L$_{\rm FIR}$/L$_{\rm B}$ $\times$ L$_{\rm FIR}$/M$_{\rm H_2}$  & 0.21 & 0.36 \\
T$_{\rm D}$ $\times$ L$_{\rm FIR}$/M$_{\rm H_2}$  & 0.63 & 0.27 \\
T$_{\rm D}$ $\times$ M$_{\rm H_2}$/M$_{\rm D}$ & 0.56 & 0.44 \\
L$_{\rm FIR}$/L$_{\rm B}$ $\times$ L$_{\rm FIR}$/(M$_{\rm H_2}$+M$_{\rm HI})$ & 0.73 & 0.46 \\
\hline
\end{tabular}
\end{table}

\begin{table}
\caption[]{Statistical Values}
\label{stat}
\begin{tabular}{lcccrc}
\hline
\multicolumn{1}{c}{Diagnostic} &
\multicolumn{1}{c}{KS} &
\multicolumn{1}{c}{KS$_{\rm Pb}$} &
\multicolumn{1}{c}{T} &
\multicolumn{1}{c}{T$_{\rm Pb}$}\\
\hline
Morphology & 0.20 & 0.74 & -0.52 & 0.60\\
log (M$_{\rm H_2}$/L$_{\rm B}$) & 0.33  &  0.16 &-1.89 & 0.07\\
log (L$_{\rm FIR}$/L$_{\rm B}$) & 0.28  &  0.33 &-1.43 & 0.16\\
log (L$_{\rm FIR}$/M$_{\rm H_2}$)&0.23  &  0.57 & 0.85 & 0.40\\
log (M$_{\rm HI}$/L$_{\rm B}$)  & 0.31  &  0.42 & 0.45 & 0.65\\
log (M$_{\rm H_2}$+M$_{\rm HI}$/L$_{\rm B}$)&0.42 & 0.09 & -1.19 & 0.24\\
log (L$_{\rm FIR}$/M$_{\rm H_2}$+M$_{\rm HI})$&0.27 & 0.55 & 0.16 & 0.87\\
log (M$_{\rm HI}$/M$_{\rm H_2}$)& 0.40  &  0.12 &  0.87 & 0.39 \\
T$_{\rm D}$   & 0.28 & 0.33 & -0.63 & 0.54\\
\hline
\end{tabular}

\noindent KS is the greatest distance between two cumulative distributions
in the Kolmogorov-Smirnov statistics. KS$_{\rm Pb}$ is the significance level to the
null hypothesis that the data sets are drawn from the same distibution. T is the Student t
coefficient for unequal variances and T$_{\rm Pb}$ is the significance level. 

\end{table}

\subsection{Nuclear Activity}

\begin{figure}
\centering
\includegraphics[width=8.8cm]{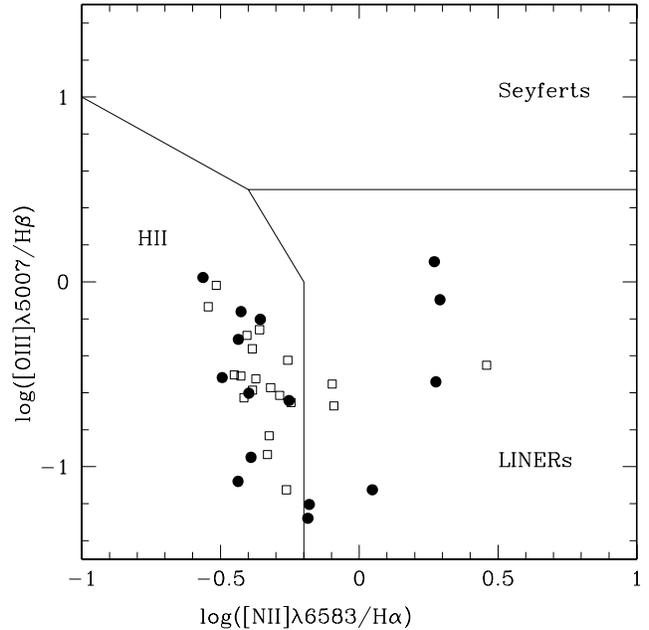}
\caption[]{Diagnostic diagram - log([OIII]$\lambda$5007/H$\beta$) 
versus log([NII]$\lambda$6583/H$\alpha$). Solid lines divide the nuclear 
activities in Seyferts, LINERs, HII galaxies, based on Veilleux \& Osterbrock 
(1987). The CS is marked by open squares. The HDS is marked by solid circles.}
\label{MS10623f3}
\end{figure}

It is believed that one of the environmental effects in disk systems is the 
efficient transport of gas to the centers of the galaxies caused by gravitational
interaction. This process can in 
principle trigger nuclear thermal (starbursts) and nonthermal activities (AGNs). 
The classification of the type of activity in our sample was done by measuring 
line-intensity ratios and applying standard diagnostic diagrams (Baldwin et al. 1981, 
Veilleux \& Osterbrock 1987). 
Fluxes have been corrected for galactic and internal 
reddening. However, due to the close wavelength separation of the lines used 
in the ratios, the internal reddening correction is nearly negligible.
In Fig.~\ref{MS10623f3} we show the log([OIII]~$\lambda$~5007/H$\beta$) versus 
log([NII]~$\lambda$~6583/H$\alpha$) for 35 galaxies (15 in the HDS and 20 in the CS). 
Most of the galaxies have spectra showing signs of star formation (HII-type).
A total of 9 LINERs were identified, 3 in the CS and 6 in the HDS. 
Field spiral galaxies and close pairs of galaxies are
reported in the literature to have 19\% and 10\% of LINERs (Ho et al. 1997 and 
Barton et al. 2000). 

The lack of Seyfert galaxies in our subsample seems to be in disagreement with the recent 
work by Coziol et al. (2000) who reported a high number of Seyfert 
galaxies in compact groups of galaxies. However, whereas our HDS subsample includes 
only intermeditate spiral galaxies, their sample includes both early- and late-type galaxies, 
with the Seyferts being more common in earlier types (E and S0s). However, intermediate type
spirals can also host AGNs. For example, in the work by Maiolino et al. (1997) where they
studied the molecular gas content of 94 Seyfert galaxies, there are 43 spirals of intermediate 
types. Therefore, the lack of Seyferts in our subsamples could be either due to a bias in 
our classification of activities or to the size of our small subsample.
The fact that we used H$\alpha$/H$\beta$=2.86 in some galaxies where it was not possible to measure
H$\beta$ in emission will not change the the diagnostic diagram significantly.
For instance, if we use H$\alpha$/H$\beta$=3.1 (typical of AGNs) instead of 
H$\alpha$/H$\beta$=2.86, the change in the vertical axis will not be large 
enough to bring the data points into the Seyfert regions. Therefore, the lack of Seyferts in our
subsample can be explained by our morphology selection and by the small number of galaxies
in or subsample. 

\begin{figure}
\centering
\includegraphics[width=8.8cm]{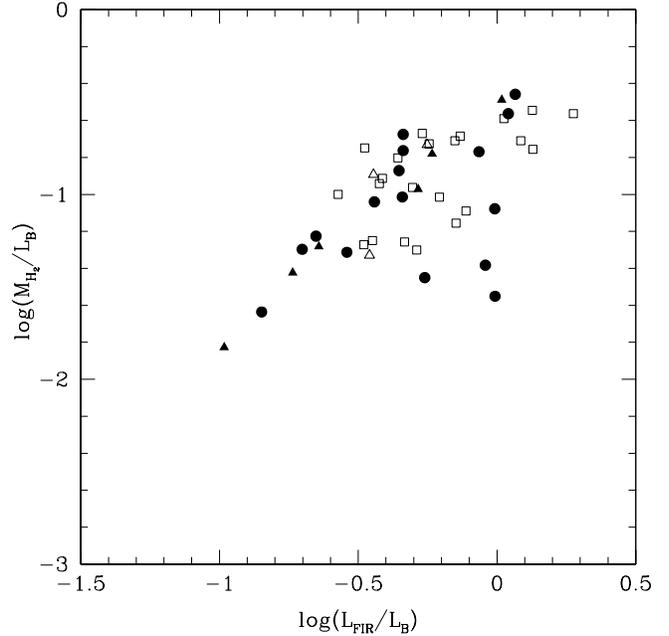}
\caption[]{Molecular gas normalized by blue luminosity as a function of 
FIR luminosity normalized by blue luminosity. The CS is marked by open symbols and HDS by 
filled symbols. LINERs are marked by triangles. Luminosities are in L$_{\odot}$ and mass 
in M$_{\odot}$.}
\label{MS10623f4}
\end{figure}

Fig.~\ref{MS10623f4} shows the correlation between the total molecular gas, M$_{\rm H_{2}}$,  
and L$_{\rm FIR}$, both divided by L$_{\rm B}$. The correlation coefficient is 0.65 and
0.59 for the HDS and CS. However, if we consider only LINERs, the correlation coefficient
increases to 0.99 and 0.75; i.e. the star formation rate (SFR) per blue luminosity 
increases linearly as the total amount of molecular gas increases in LINERs, 
in particular for the HDS. This result might have important implications in the 
understanding of the nature of LINERs (e.g. Alonso-Herrero et al. 2000).
However, a larger sample of LINERs should be used in order to statistically test the 
significance of this trend. 
Another interesting result is the fact that the galaxies which deviates from the 
linear correlation are all non-AGN galaxies. Hence, AGN heating of dust cannot 
be invoked as an explanation for the higher L$_{\rm FIR}$/L$_{\rm B}$ for a given 
M$_{\rm H_{2}}$/L$_{\rm B}$.
 
\subsection{Dust temperature}

Dust temperature can provide a better understanding of the physical conditions 
inside the galaxies.  Warm dust (T$_{\rm D}$ $>$ 50 K) are typical of molecular clouds 
where massive stars ($>$ 6 M$_{\odot}$) reside, whereas cold dust 
(T$_{\rm D}$ $<$ 30 K) trace quiescent molecular clouds heated by the
interstellar radiation field. However, the limited spatial resolution of IRAS
gives an average of extended cold dust emission and small hot emission areas
(the dust emissivity goes as $T_{\rm D}^4$). Nevertheless, a higher dust temperature is
indicative of current star formation. From the ratio of the IRAS fluxes at 60 $\mu$m and
100 $\mu$m we derived dust temperatures assuming $\kappa$$_{\nu}$ $\propto$ $\nu$.
Dust masses were derived using the following equation,

M$_{\rm D}$ = 4.8 $\times$ 10$^{-11}$ $\alpha$ S$_{100}$ d$^{2}$/($\kappa$$_{\nu}$B$_{\nu}$
(T$_{\rm D}$))

\noindent where S$_{100}$ is the IRAS flux at 100 $\mu$m in Jy, $\kappa$$_{\nu}$ is the mass
opacity of the dust ($\kappa$$_{\nu}$=25 cm$^{2}$g$^{-1}$, Hildebrand 1983), 
B$_{\nu}$(T$_{\rm D}$) is the Planck function, d is the distance in Mpc, and $\alpha$ the
molecular gas-to-dust mass ratio ($\alpha$=700, Thronson \& Telesco 1986). 

\begin{figure}
\centering
\includegraphics[width=8.8cm]{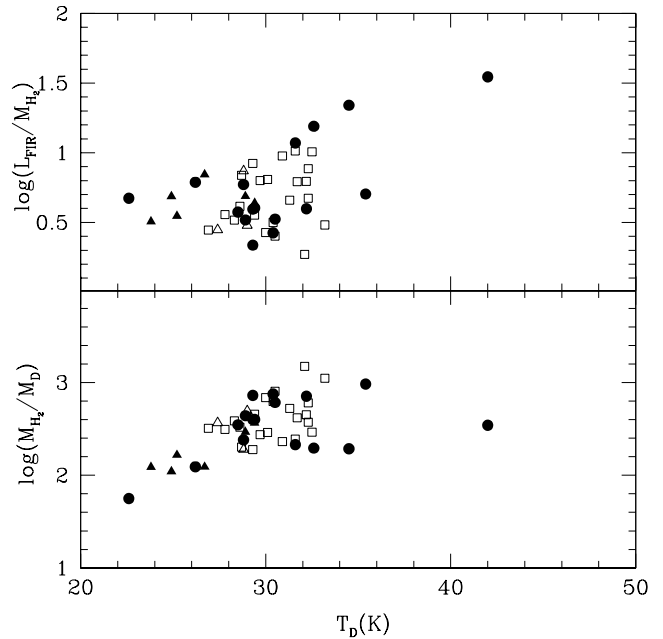}
\caption[]{Upper panel: FIR luminosity normalized by
molecular gas as a function of
dust temperature in K. Lower panel: molecular gas normalized by
dust mass as a function of dust temperature in K. Luminosities are in L$_{\odot}$ and mass in 
M$_{\odot}$. Symbols are the same as in Fig.~\ref{MS10623f4}.}
\label{MS10623f5}
\end{figure}

The average values of T$_{\rm D}$ (HDS=31.3 $\pm$ 2.8 and CS=30.1 $\pm$ 2.4) are typical of 
spiral galaxies (Sage 1993, Wiklind et al. 1995). In Fig.~\ref{MS10623f5} we see that there is a lack of correlation between 
SFE and T$_{\rm D}$ for the CS and a weak correlation for the HDS (correlation 
coefficients are $\sim$ 0.3 and 0.6, respectively).
However, the relatively low values of T$_{\rm D}$ for 
LINERs (Table~\ref{liners}) might have important implications 
regarding the interpretation of the source that powers the nuclear region 
of these galaxies. Low T$_{\rm D}$ implies a low current star formation and no 
powerful black hole. 
The scenario proposed by Alonso-Herrero et al. (2000) fits well our results. 
If LINERs are aging starbursts they should have low T$_{\rm D}$ since their 
massive stars would have evolved after 5--10 Myr. This is also suggested by
the lower averages of log L$_{\rm FIR}$/M$_{\rm H_2}$ for LINERs. 
However, we cannot exclude 
the presence of a central black hole with reduced activity (Ji et al. 2000). 
In view of the small number of LINERs in our sample it is not possible to 
reach any firm conclusion regarding the correlation between the amount of 
fueling gas and the starburst and/or AGN activities.

\begin{table}
\caption[]{T$_{\rm D}$ and SFE average values}
\scriptsize
\label{liners}
\begin{tabular}{c|cc|cc}
\hline
\multicolumn{1}{c}{} &
\multicolumn{2}{c}{LINERs} &
\multicolumn{2}{c}{Non-LINERs} \\
\multicolumn{1}{c}{Diagnostic} &
\multicolumn{1}{c}{HDS} &
\multicolumn{1}{c}{CS} &
\multicolumn{1}{c}{HDS} &
\multicolumn{1}{c}{CS} \\
\hline
T$_{\rm D}$ & 26.5$\pm$2.3 & 28.4$\pm$0.8 & 31.3 $\pm$2.8 & 30.1 $\pm$2.4\\ 
log L$_{\rm FIR}$/M$_{\rm H_{2}}$ & 0.65$\pm$0.12 & 0.60$\pm$0.23 & 0.77$\pm$0.35 & 0.68$\pm$0.21\\
\hline
\end{tabular}
\end{table}  

\subsection{Star Formation Rate}

The H${\alpha}$ equivalent width (EW) is defined as the emission-line luminosity normalized to the
adjacent continuum flux, and hence is proportional to the star formation rate per unit (red) luminosity.
The mean value of EW(H${\alpha}$) for the HDS and CS are 14.9 $\pm$ 11.7 and 13.4 $\pm$ 9.5 \AA, respectively. 
These values are lower than the ones found in Pastoriza et al. 
(1994) for galaxies in other subsamples of HDS and CS galaxies. The main difference between the latter 
subsample and our subsample is the morphology and type of activity. The majority of the galaxies 
in Pastoriza et al. (1994) with large EWs are either later type spirals or classical Seyfert 
galaxies, like NGC7582. In our subsample we have spirals of intermediate type and no Seyferts.

Since the HDS has more LINERs than the CS and LINERs tend to have smaller
EW(H${\alpha}$), we find that, if we disregard all LINERs, the mean value of EW(H${\alpha}$)
changes to 19.8 $\pm$ 10.1 and 15.3 $\pm$ 9.0\AA, for the HDS and CS, respectively.
However, the number of galaxies with measured EW(H${\alpha}$) is too small to give any 
statistically significant result. SFRs based on L(H${\alpha}$) can also be severely
underestimated due to dust internal to the galaxies (Bushouse 1987, Kennicutt 1998). Therefore, 
we use L$_{\rm FIR}$ as a diagnostic to the recent star formation.


\section{Discussion}

\subsection{Environment}

\begin{figure}
\centering
\includegraphics[width=8.8cm]{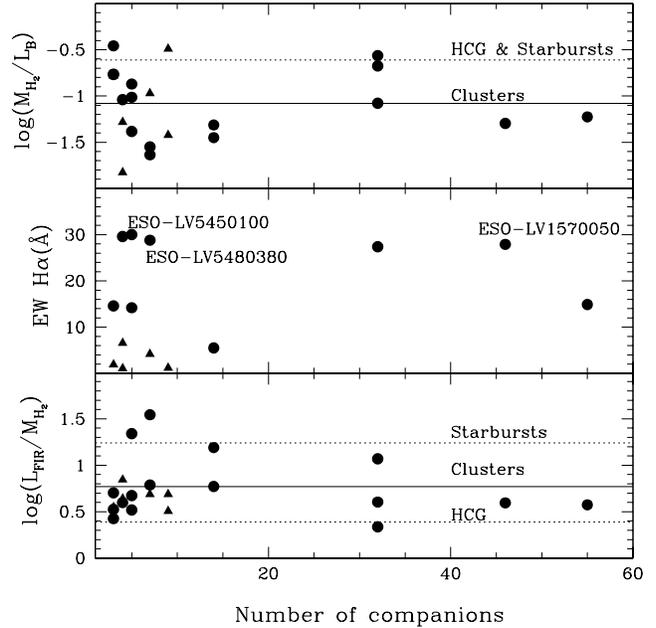}
\caption[]{Upper panel: Molecular gas normalized by 
blue luminosity as a function of the number of companions of galaxies (N$_{\rm g}$) in the HDS. 
Middle panel: the equivalent width of H${\alpha}$ in \AA\ as a function of N$_{\rm g}$
in the HDS. 
Lower panel: FIR luminosity normalized by molecular gas as a function of N$_{\rm g}$
in the HDS. Horizontal lines are average values for Hickson compact groups (HCG), 
starbursts and clusters from Leon et al. (1998). LINERs are marked by triangles. 
Luminosities are in L$_{\odot}$ and mass in M$_{\odot}$.}
\label{MS10623f6}
\end{figure}

\begin{figure}
\centering
\includegraphics[width=8.8cm]{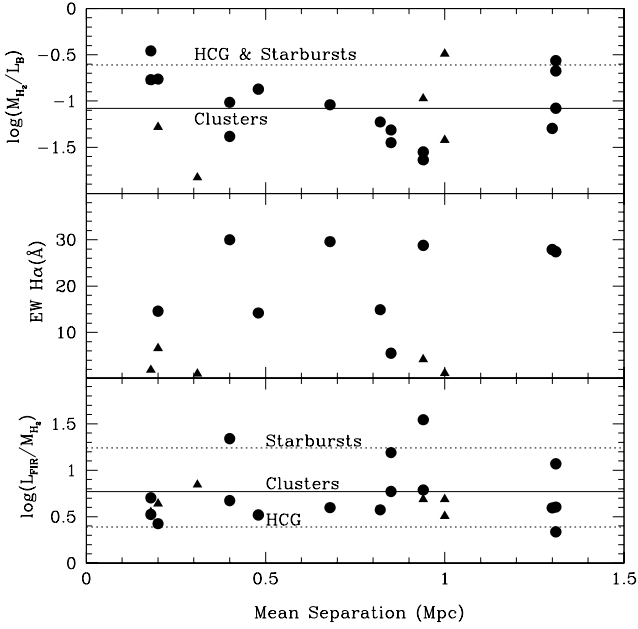}
\caption[]{Upper panel: Molecular gas normalized by blue luminosity as a function of 
the mean separation (r$_{\rm p}$) between galaxies in the HDS.
Middle panel: the equivalent width of H${\alpha}$ in \AA\ as a function of 
r$_{\rm p}$ in the HDS.
Lower panel: FIR luminosity normalized by molecular gas as a function of 
r$_{\rm p}$ in the HDS. Horizontal lines are for 
Hickson compact groups (HCG), starbursts and
clusters from Leon et al. (1998). LINERs are marked by triangles. 
Luminosities are in L$_{\odot}$, mass in M$_{\odot}$, and r$_{\rm p}$ in Mpc.}
\label{MS10623f7}
\end{figure}

In order to investigate whether there is any correlation between the environment, the
type of activity, and the total amount of molecular gas, we have identified galaxies in 
the HDS which are also part of groups according to Maia et al. (1989).
As environmental parameters we used the number of companions and the mean separation 
between the galaxies of each group (N$_{\rm g}$ and r$_{\rm p}$ in Maia et al. 1989).
We plotted M$_{\rm H_2}$/L$_{\rm B}$, EW(H${\alpha}$) and SFE, as a 
function of N$_{\rm g}$ and r$_{\rm p}$ (Fig.~\ref{MS10623f6}, Fig.~\ref{MS10623f7}). 
We also marked in these figures the mean values for clusters of galaxies, starburst 
galaxies, and Hickson compact groups. No correlation between M$_{\rm H_2}$/L$_{\rm B}$ and
the environmental parameters is found. There are galaxies 
as gaseous as starbursts and as poor as clusters in all environments. 
LINERs were found only in groups with less 
than 20 members but are present in groups with either small or large separations 
between the members.

We also found no correlation between EW(H${\alpha}$) and the environmental parameters. 
This is in disagreement with Barton et al. (2000) who found 
that EW(H${\alpha}$) anticorrelates strongly with pairs spatial separation and 
velocity separation. However, this could be either due to the small size of our 
sample in comparison with Barton et al. (2000) which contains 502 galaxies, 
and/or to the differences in the selection criterion imposed by our sample 
and Barton et al.  The latter sample mixes pairs and groups of galaxies 
whereas our sample has galaxies in dense environments and no isolated pairs of
galaxies.

The lack of a trend in Fig.~\ref{MS10623f6} and Fig.~\ref{MS10623f7} is clear. However, we 
decided to check some galaxies individualy.
We noticed that three of the HDS galaxies (ESO-LV1570050, ESO-LV5450100, and ESO-LV5480380) with the 
largest values of EW(H${\alpha}$) have less M$_{\rm H_2}$/L$_{\rm B}$ than the 
average value.
They are located in groups with small and large separations 
(e.g. r$_{p}$$_{\rm ESO-LV5450100}$=0.40 Mpc, r$_{p}$$_{\rm ESO-LV5480380}$=0.94 Mpc, 
and r$_{p}$$_{\rm ESO-LV1570050}$ = 1.3 Mpc) and with small and large number of companions 
(ESO-LV5450100 has only 4 companions, ESO-LV5480380 has 6 companions whereas ESO-LV1570050 has 
45 companions). Therefore, 
for these three galaxies, these environmental parameters are not responsible for 
the enhancement of star-formation and defficiency in molecular gas.

\begin{figure}
\centering
\includegraphics[width=8.8cm]{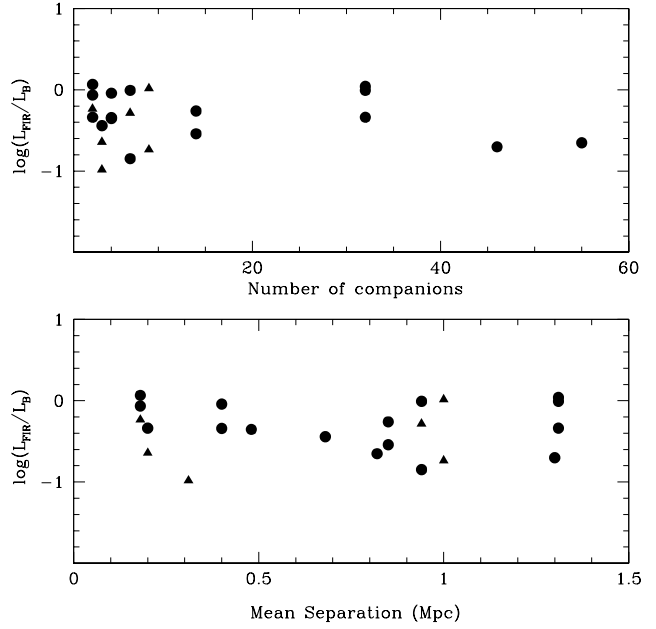}
\caption[]{Upper panel: Number of companions (N$_{\rm g}$) of galaxies in the HDS as a function of FIR luminosity 
normalized by the blue luminosity. Lower panel: Mean separation (r$_{\rm p}$) between the members of the groups 
in the HDS as a function of FIR luminosity normalized by the blue luminosity. 
LINERs are marked by triangles. Luminosities are in L$_{\odot}$ and r$_{\rm p}$ in Mpc.}
\label{MS10623f8}
\end{figure}

There is also no correlation between SFE and the environmental parameters.
(Fig.~\ref{MS10623f6} and Fig.~\ref{MS10623f7} lower panels). Combes et al. (1994) also 
found no correlation between SFE and separation in pairs of galaxies. However, because 
pairs of galaxies are FIR enhanced they suggested that pairs have an increase in the 
molecular gas content. However, this is not the case for our sample. No clear 
trend between the environment and L$_{\rm FIR}$/L$_{\rm B}$ is found (Fig.~\ref{MS10623f8}). 
We note that the two galaxies with the largest number of companions have the 
lowest L$_{\rm FIR}$/L$_{\rm B}$. This could be an indication of the influence of the
environment on the star formation rate of these galaxies and should be investigated for
a larger sample. Similar results were suggested by Hashimoto et al. (1998) using the [OII]
emission line as a diagnostic of SFR. 

The possibility of a correlation with the crossing time 
($t_{v}$ in Maia et al. 1989) was also checked and no correlation was found. 
Therefore, we conclude that {\it there is no clear correlation between molecular gas content, SFR, SFE 
and the environmental parameters}.

\subsection{Total Gas and Morphology}

\begin{figure}
\centering
\includegraphics[width=8.8cm]{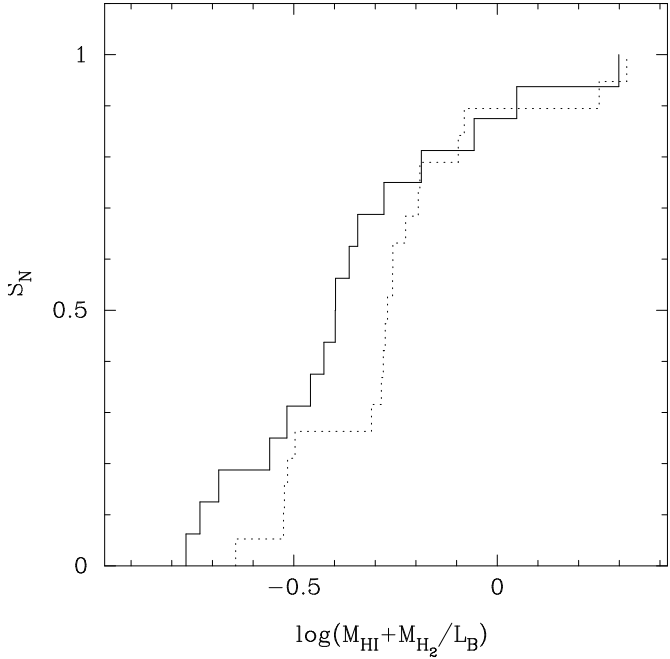}
\caption[]{Cumulative probability distribution of the total gas normalized by 
total gas. HDS is marked by solid line and CS by dotted line. Luminosity is  
in L$_{\odot}$ and mass in M$_{\odot}$.}
\label{MS10623f9}
\end{figure}

\begin{figure}
\centering
\includegraphics[width=8.8cm]{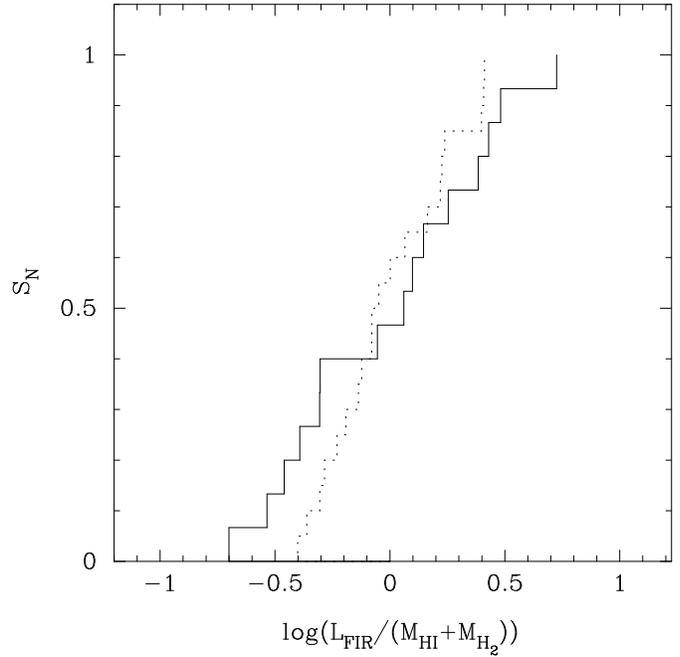}
\caption[]{Cumulative probability distribution of FIR luminosity normalized by 
total gas. HDS is marked by solid line and CS by dotted line. Luminosity is  
in L$_{\odot}$ and mass in M$_{\odot}$.}
\label{MS10623f10}
\end{figure}

\begin{figure}
\centering
\includegraphics[width=8.8cm]{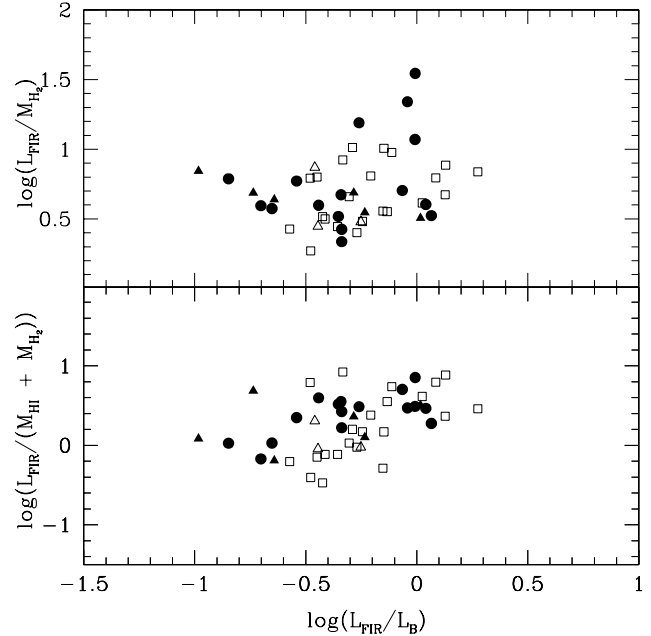}
\caption[]{Upper panel: FIR
luminosity normalized by molecular gas as a function of FIR luminosity normalized by blue luminosity
Lower panel: FIR luminosity normalized by total (neutral and molecular) gas as a function of 
FIR luminosity normalized by blue 
luminosity. Symbols are the same as in Fig.~\ref{MS10623f4}. Luminosities are in L$_{\odot}$ and mass in 
M$_{\odot}$.}
\label{MS10623f11}
\end{figure}

\begin{figure}
\centering
\includegraphics[width=8.8cm]{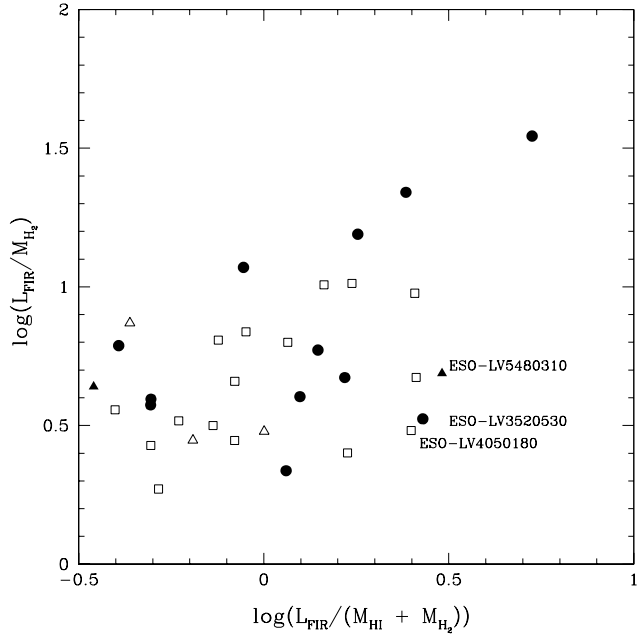}
\caption[]{FIR luminosity normalized by molecular gas as a 
function of FIR luminosity normalized by total (atomic and molecular) gas. 
Symbols are the same 
as in Fig.~\ref{MS10623f4}. Luminosities are in L$_{\odot}$ and mass in 
M$_{\odot}$.}
\label{MS10623f12}
\end{figure}

\begin{figure}
\centering
\includegraphics[width=8.8cm]{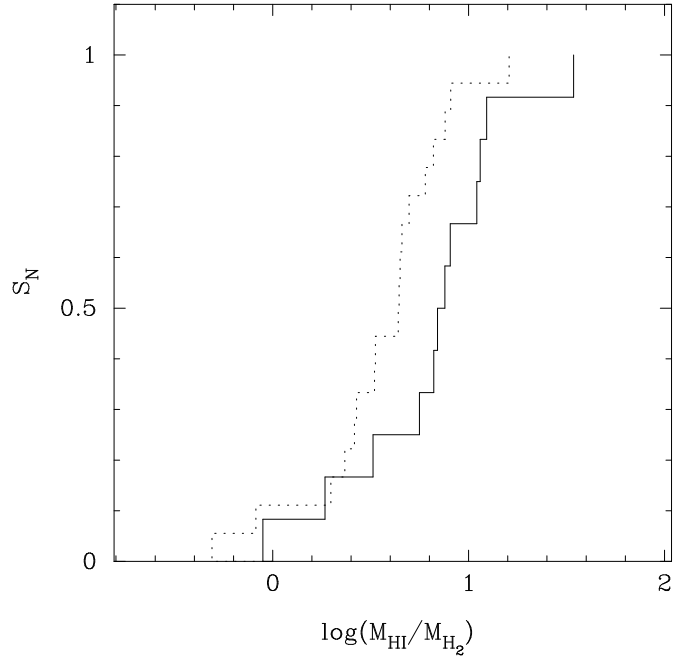}
\caption[]{Cumulative probability distribution of atomic and molecular gas ratio. 
HDS is marked by solid line and CS by dotted line.}
\label{MS10623f13}
\end{figure}

It is possible that interacting galaxies like galaxies in compact groups 
have less current star formation because they have less 
total fuel (Sulentic \& de Mello Raba\c ca 1993, Leon et al. 1998).  Therefore, 
the total amount of gas, H$_{2}$~+~HI, might be a better SFE indicator in very dense environments. 
In order to check if this is valid for our sample, we used the HI data available in the NED for
35 galaxies of our sample. Our results are: 

\begin{itemize}

\item {\bf Total gas}: (M$_{\rm HI}$+M$_{\rm H_{2}}$)/L$_{\rm B}$ distributions (Fig.~\ref{MS10623f9}) are significantly 
different (91\% level).

{\it The HDS has lower total gas than the CS}.

\item {\bf Gas Fraction}: the average values of log (M$_{\rm HI}$/M$_{\rm H_{2}}$) for the HDS and CS are 
0.69$\pm$0.59 and 0.51$\pm$0.46. M$_{\rm HI}$/M$_{\rm H_{2}}$ distributions are 
significantly different (88\% level).

{\it The HDS in comparison with CS 
have, more atomic gas, or higher atomic gas 
fraction}. Similar values (log (M$_{\rm HI}$/M$_{\rm H_{2}}$)=0.62$\pm$0.43) are found by 
Horellou \& Booth (1997) for a sample of southern interacting galaxies.

\item {\bf SFE (L$_{\rm FIR}$/total gas)}: the star formation efficiency calculated using the total gas, 
L$_{\rm FIR}$/(M$_{\rm HI}$+M$_{\rm H_{2}}$), distributions (Fig.~\ref{MS10623f10}) are not 
significantly different (55\% level). 

Fig.~\ref{MS10623f11} shows the SFE indicators as a function 
of L$_{\rm FIR}$/L$_{\rm B}$ (both plots include only galaxies for which HI data 
were available). The large dispersion seen in the upper panel decreases in the lower
panel where we include the neutral gas. The correlation coefficient changes from
0.21 to 0.73 in the HDS and 0.59 to 0.46 in the CS (r in Table~\ref{coeffs}). 

Fig.~\ref{MS10623f12} shows the SFE including the atomic gas, 
L$_{\rm FIR}$/(M$_{\rm HI}$+M$_{\rm H_{2}})$, as a function of the SFE without 
the atomic gas. The three most efficient galaxies previously found using only 
the molecular gas (ESO-LV5480380, ESO-LV5450100, ESO-LV1190060) still have high values of SFE  
when we include the HI content. However, it is interesting to see how other galaxies 
(ESO-LV3520530, ESO-LV4050180, and ESO-LV5480310 - marked in Fig.~\ref{MS10623f12}) 
which were not among the efficient ones, now have 
higher values. This could be due to the low amount of HI in these 
{\it earlier types spirals}. If we remove them from our sample and keep a
more homogeneous sample in terms of morphology, we find that 
{\it log (M$_{\rm HI}$/M$_{\rm H_{2}}$) is higher for the HDS than for the CS}. The
cumulative distribution function (Fig.~\ref{MS10623f13}) of the HDS is significantly 
different than that of the CS. However, we have HI data for only 12 galaxies of the HDS 
which are later than Sb, therefore, a larger and homogeneous sample in terms of morphology 
should be observed in order to confirm this result.

\end{itemize}

\section{Summary and Conclusions}

We have performed a detailed comparison between properties of intermediate Hubble type 
galaxies in dense environments (HDS) and in the field (CS). By using several different
diagnostics of global properties we have found a trend  for the gaseous content
and star formation properties of the high and low density samples. Intermediate Hubble type galaxies 
in dense environments have, on average:

\begin{itemize}

\item {\bf lower} gas content than field galaxies 
(i.e. lower M$_{\rm gas}$/L$_{\rm B}$ ratio)

\item {\bf higher} atomic gas fraction than
field galaxies (i.e. a higher M$_{\rm HI}$/M$_{\rm H_2}$ ratio)

\item {\bf lower} current star formation rate
than field galaxies (i.e. lower L$_{\rm FIR}$/L$_{\rm B}$ ratio)

\item the {\bf same} star formation efficiency as field galaxies 
(i.e. the same L$_{\rm FIR}$/M$_{\rm H_2}$ or L$_{\rm FIR}$/M$_{\rm gas}$
ratio)

\end{itemize}

Although none of the above results stand out as a single strong diagnostic given their
statistical significance (see Table 3), taken together they suggest a trend for
diminished gas content and star formation activity in galaxies in high density environments.

What can be the physical processes behind this result? It has long been believed that
gravitational interaction is a sufficient condition for transporting gas from the outer
regions of galaxies to the inner regions, and thereby increasing the star formation activity.
If this is the case for our high density sample, the lower gas content could be the result
of gas exhaustion through an increased star formation activity {\it in the past}.
However, it seems unlikely that we should have selected only those systems which experienced
enhanced star formation rates in the past. An alternative and more likely interpretation is
that repeated close encounters, experienced by galaxies in dense environments, removes gas
from the galaxies as well as leads to an inhibition of the formation of molecular gas from
the atomic phase.
The similarities in star formation efficiency then suggest that the physical processes
controling the formation of stars from the molecular gas are local rather than global.

\medskip

In addition, we find that 6 (38\%) of the HDS galaxies and 3 (14\%) of the CS galaxies are
classified as LINERS. These are very limited numbers but a few interesting trends among
the LINERS are suggested by our data.
First of all, the LINERS show a very good correlation between molecular gas and L$_{\rm FIR}$,
they have a lower dust temperature than the non-LINER HDS and CS galaxies and they have
a lower average star formation efficiency. These findings are consistent with the LINERS
as aging starbursts rather than being powered by an AGN. A larger sample is needed to
study these correlations in more detail.


\begin{acknowledgements}
The ON team of observers at the ESO1.52m, in particular Christopher Willmer for helping
with the data reduction. Henrique Schmitt for valuable suggestions regarding the stellar contamination.
To the anonymous referee for valuable suggestions which helped improving our paper. 
This research has made use of the NASA/IPAC Extragalactic Database (NED) which is
operated by the Jet Propulsion Laboratory, California Institute of Technology, under contract with the
National Aeronautics and Space Administration. D.F.M. was supported partially by CNPq Fellowship
301456/95-0, and the Swedish 
\emph{Vetenskapsr{\aa}det} (NFR) project number F620-489/2000. M.A.G.M. was supported by
CNPq grant 301366/86-1. T.W. was supported by Vetenskapsr{\aa}det project number F1299/1999.
\end{acknowledgements}
\end{document}